\documentclass[10pt]{book}
\def\shorttitle{The Heart of Matter}
\usepackage{insa}
\usepackage{amsmath,amssymb}
\usepackage{graphicx}% Include figure files
\usepackage{epsfig}% Include figure files
\usepackage{rotating}% Include figure files
\usepackage{dcolumn}% Align table columns on decimal point
\usepackage{color}
\usepackage{pstricks,pst-node,pst-text,pst-3d}
\usepackage{bm}% bold math

\def\shorttitle{The Heart of Matter}
\begin{document}

\title{The Heart of Matter}

\author{Rohini M. Godbole}%{FNA, FNASc, FASc}

\affil{Theoretical Physics, CERN, CH-1211, Geneva-23,Switzerland\\
Center for High Energy Physics, Indian Institute of Science, Bangalore 560 012, India}

\beginabstract

In this article I trace the development of the human understanding of  the
``Heart of Matter'' from  early concepts of ``elements'' (or alternatively
``Panchmahabhootas'') to the current status of ``quarks'' and ``leptons'' as the
fundamental constituents of matter, interacting together via exchange of the
various force carrier particles called ``gauge bosons'' such as the photon, 
W/Z-boson etc. I would like to show how our understanding of the 
fundamental constituents of matter has gone hand in hand with our 
understanding of the fundamental forces in nature.  I will 
also outline how the knowledge of particle physics at the ``micro'' scale of
less than a Fermi(one millionth of a nanometer), enables us to 
offer explanations of Cosmological observations at the ``macro'' scale. 
Consequently these observations, may in turn, help us address some very 
fundamental questions of the Physics at the ``Heart of the Matter''.

\endabstract

\section*{1. Concept of ``elementarity'' through ages.}
In addressing any problem in any walk of life, the recognition of the central
issue is always essential. A query of what lies at the 
``heart'' of a given problem is of utmost importance to all of us, in dealing 
with various issues in everyday life. It is therefore, not surprising that,
since the dawn of humanity, a major part of the scientific endeavor of the 
humankind has been devoted to gain an understanding as to what lies at the
`Heart of the Matter'. The scientific knowledge and process as we know today, 
has developed through a desire to know how  nature operates. One of the
central themes in these explorations has been the wish to know whether all the 
matter is made up of elemental building blocks and if so  how these elemental 
constituents are held together. In more colloquial words one might call this
a quest for deciphering what the  ``bricks'' and ''mortar' of this  wonderful
edifice of life around us are. Funnily enough through the ages, the development
of our understanding of what the fundamental constituents of matter are, has 
grown hand in hand along with our knowledge of the working of various 
fundamental processes and the fundamental forces of nature. It is
this interplay that I find most fascinating.

At present particle physicists have arrived at an understanding of the basic 
laws of physics which govern the behaviour of the fundamental building blocks
of matter, the quarks and the leptons. The interesting fact is that the same
laws, {\Large \it in principle}, allow us to predict the behavior of all the matter 
around us under all circumstances.
Indeed we have come a long way since the early days of the Greek Empedocles 
who thought that the world was made up of the four {\it elements}: 
Earth, Water, Fire, Air. So also from the days of the early Indian sages 
who identified the five entities: the above four along with the ``sky'' as
the {\bf Panchmahabhootas}, as those whose workings need to be understood 
and which need to be conquered. Starting with this really ``small'' number 
of fundamental
``elements'' of nature, our concept of elementarity has evolved through the 
ages, starting from  molecules, atoms, nuclei and finally ending in 
quarks/leptons  after passing through protons/neutrons on the way,
as the candidates for the basic building blocks of matter. Finally, 
it seems to have come home,
at least temporarily,  to roost in the wonderful picturesque world 
of ``elementary particle physics''. The particle physicists, for good reason, 
believe that now we have perhaps peeled the last layer of the onion and 
the nature has revealed the ultimate constituents of matter to us. We feel
that we have seen the last faceless entity at the heart of this Russian Doll. 
The currently accepted list of the elementary particles consists of the quarks,
leptons, the force carrier particles called gauge bosons: $\gamma, W^\pm, 
Z^0$, gluons along with  the as yet {\it undiscovered} Higgs boson. It is this 
journey starting from the ``elements'' of the early Greeks/Indians to the 
quarks/leptons as the fundamental constituents, that I want to sketch out 
for you. 

The subject of elementary particle physics, which is the branch of physics
that deals with the ultimate layer of structure of matter, addresses  
the following three issues. These are:\\
$\bullet$ What are the elementary constituents of matter? \\
$\bullet$ What holds them together? \\
$\bullet$  What is the correct mathematical framework to describe how the
constituents are put together to form matter, how do they interact with each
other and how can one predict its behavior under different conditions? \\

%\begin{itemize}
%\item[1] What are the elementary constituents of matter?
%\item[2] What holds them together?
%\item[3] What is the correct mathematical framework to describe how the 
%constituents are put together to form matter, how do they interact with each 
%other and how can one predict its behavior under different conditions? 
%\end{itemize}

Interesting thing is that the path to the correct answer to the first two 
questions at a given level of elementarity, has been indicated only by the 
answers to the last question at the earlier level of elementarity. 
A detailed account of the aspect of the elementary particle 
physics mentioned under point (3) above, is to be found in the
article of Prof. A. Raychaudhuri, elsewhere in this volume. 
I would therefore not really spend much time on it, rather, I would 
like to chart out 
for you how our ideas of elementarity have changed and why we believe that 
quarks and leptons are indeed the {\bf `fundamental'} constituents. This means
I will not  discuss much about the ``force carriers''. One basic point I want 
to make is, that in the end the essential process by which structure 
has been revealed has been more or less the same at all levels.

Equally interesting is another development of the past few decades, which have
made us realise that this world at the smallest distance scales\footnote
{Indeed the size of an electron, if it all it is not a 'point', has to be less 
than a million, million, million$^{\rm th}$ of a meter stick} holds clues to 
some of the puzzles of the Cosmos with its huge distance scales of millions of 
parsecs\footnote{One parsec is roughly 180 times the earth-sun distance, {\it 
i.e.} 27 Billion Kilometers.}, such as why matter dominates over antimatter in 
the Universe or what might be the ``dark matter'' which does not shine but
whose existence is revealed through its gravitational effects etc. The results 
of on going investigations in different High Energy Physics experiments at 
the colliders or otherwise should be able to confirm whether the explanations 
offered by the HEP theory to these puzzles are indeed the correct ones. 
Currently the most important puzzle of them all is why our Universe is 
accelerating?  A new development in  Particle physics theory extended to 
include gravitation, called the String theory, might have a solution for 
that as well! This interplay between the ''micro'' and the ''macro'' scale is
one of the most amazing things and reminds me of a saying by Albert
Einstein, which I freely paraphrase: `The most incomprehensible thing about 
our Universe is that it is comprehensible to human thought'.

\section*{2. Standard Model of Particle Physics}
Let me begin by summarising the currently accepted picture of the fundamental 
constituents and interactions among them, the Standard Model of Particle 
Physics(SM), before I venture into a retracing of the tortuous path taken by 
the scientific community from the time of the Demorkritos and Kanad to arrive 
at the SM. According to our current understanding, not only the bricks but 
even the mortar (the force carriers) are elementary particles.  
\begin{figure}[htb]
\begin{center}
\includegraphics*[width=12cm]{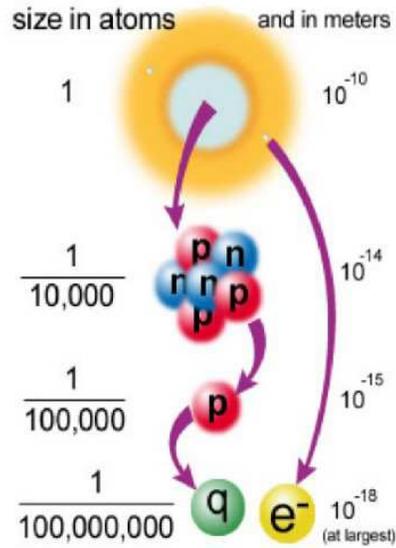}
\caption{\it Constituents of matter at different distance scales}
\label{figscale}
\end{center}
\end{figure}
Fig.~\ref{figscale}
shows the constituents of matter at different distance scales, beginning from 
atoms with a size of one tenth of a billionth meter, ending with {\it quarks} 
and {\it leptons}.  Experiments put an upper limit on their size, which itself 
is a hundred million times smaller than the size of an atom and today are 
believed to be truly indivisible. These are considered to be the fundamental 
constituents of matter. Experiments at high energy accelerators, and the 
development of theoretical models, have together helped us arrive at this
conclusion.
\begin{table}[ht]
\begin{center}
\begin{tabular}{|c|c|}
\hline
&\\
Quarks & Leptons \\ 
&\\
\hline
&\\
$\left( \begin{array}{c} u\\ d\\ \end{array} \right)$
\hspace{0.2cm} 
$\left( \begin{array}{c} c\\ s\\ \end{array} \right)$
\hspace{0.2cm} 
$\left( \begin{array}{c} t\\ b\\ \end{array} \right)$
& 
$\left( \begin{array}{c} e^-\\ \nu_e\\ \end{array} \right)$
\hspace{0.2cm} 
$\left( \begin{array}{c} \mu^-\\ \nu_\mu\\ \end{array} \right)$
\hspace{0.2cm} 
$\left( \begin{array}{c} \tau^-\\ \nu_\tau\\ \end{array} \right)$
\\ 
$\times$ 3 colours & \\ 
 & \\
%${\red \left( \begin{array}{c} u\\ d\\ \end{array} \right)}$
%\hspace{0.2cm} 
%${\green \left( \begin{array}{c} u\\ d\\ \end{array} \right)}$
%\hspace{0.2cm} 
%${\blue \left( \begin{array}{c} u\\ d\\ \end{array} \right)}$
%& `colourless Leptons' \\ 
%{\it etc.}&\\
&\\
+anti-quarks & + anti-leptons \\ \hline
\end{tabular}
\caption{\it The fundamental constituents of matter.}
\end{center}
\end{table}

The quarks and leptons come in several different varieties, summarised
in Table 1. The quarks are called u(p), d(own), c(harm), s(trange),t(op) and 
b(ottom) whereas the leptons are the well known electron(e) along with
muon($\mu$), tau-lepton ($\tau$) and the corresponding neutrinos, 
$\nu_e, \nu_\mu$ and $\nu_\tau$ respectively. Later we 
will get a glimpse of why so many varieties must be present.

As for the forces we know today that there are of four basic forces 
experienced by the constituents of matter:
\begin{enumerate}
\item{}  Gravitational Force: The force that holds us on the earth,
and gives rise to planetary motion as well as tides.
\item{} Electromagnetic Force: The force that holds electrons inside
atoms, and that is responsible for electrostatic effects, electric
currents, and magnetic poles.
\item{} Weak Force: The force that causes the decay of radioactive nuclei, 
in which a proton changes into a neutron or vice versa.
\item{} Strong Force : The force that binds together the
quarks inside protons and neutrons, and also makes the latter stick to
each other to form the atomic nucleus.
\end{enumerate}
The force responsible for holding the nucleons (protons and neutrons)
together in a nucleus, is derived from the Strong Force above in a
similar way that the ``Van der Waals'' force (the force between neutral atoms 
holding them together in a molecule) is derived from the Coulomb interaction 
among the charged constituents of the otherwise neutral atom.
These forces are familiar to us to varying degrees, depending on their
effects on the kind of objects that we encounter in daily life. 
The first two forces in the above list have been known
almost since the dawn of scientific thought, while the last two are
nuclear forces and were discovered only in the twentieth century. The
effects of the latter two forces cannot be observed directly by the
human senses, but they are just as real as the first two, since
experimental equipment is certainly able to detect them.
For the strong, electromagnetic and weak interaction, we have been
able to show that these  interactions between the constituents of 
matter are conveyed via the mediation of the force carriers. 
The interactions along with these force carriers are listed in Table 2.
\begin{table}[ht]
\begin{center}
\begin{tabular}{|c|p{2.25in}|c|}
\hline
&&\\
Interaction&Description&Carrier Particle\\
&&\\
\hline
&&\\
Gravitation&Long-range but extremely weak attraction between all particles. & ????\\
%&between all particles.&\\
&&\\
Electromagnetic&Long-range interaction of a quark or lepton 
with another quark or lepton &Photon $\gamma$\\
%&with another quark or lepton&\\
&&\\
Weak&Short-range interaction that can cause different quarks
and leptons to change into each other &$W/Z$ Bosons\\ 
%& and leptons to change into each other&\\
&&\\
Strong&Short-range interaction among quarks only &Gluons $g$\\[3mm]
\hline
\end{tabular}
\caption{\it Four basic forces in Nature, and the carriers for three of them.}
\end{center}
\end{table}

The lighter quarks manifest themselves only as
bound states like protons, pions and kaons. The neutrinos have only weak
interactions, whereas the colourless charged leptons have weak and
electromagnetic interactions and the coloured quarks feel all the three
interactions.
The properties of all the particles, the constituent matter particles
and the force carriers, have been measured to a high degree of accuracy.

Let us recall here that most elementary particles carry a ``spin'', or
intrinsic angular momentum. We believe this because of experiments in
which particles are found to behave as if they are spinning on their
own axis. This is not literally true: if it were, one should be able
to change their amount of spin, or stop them from spinning, while in
fact their spin angular momentum is an unchangeable property. So we
must treat it as an intrinsic property of the particle. It turns out
that it has to be always an integer or half-integer multiple of a
basic unit called $\hbar$ or Planck's constant. This multiple is
called the 'spin' of the particle. Particles of integer spin are
called ``bosons'' and those of half-integer spin, ``fermions''.
All these particles have a few other intrinsic properties which have been
given very imaginative and descriptive names such as strangeness, colour
etc. by particle physicists and I will get back to that a little later.

The achievement of the
last fifty odd years of the particle physicists as a community has been to 
arrive at an understanding of the working of the matter particles, the 
quarks and leptons, and the force carriers, the bosons and develop the 
mathematical framework in which this can be described. The latest in this 
series of developments is to develop a mathematical framework which will 
also might  make it possible to describe workings of gravity at the same level. 
As already mentioned this subject does not concern us here. We will now
proceed  to discuss how particle physicists arrived at an understanding 
that these quarks and leptons are the building blocks of nature.

\section*{3. A tale of molecules, atoms and nuclei}
Our concept of elementarity has undergone a change in the centuries; 
so has the  branch of science in general and physics in  particular, that
has dealt with the issue. Demokritos said {\it 
`By convention there is colour, by convention sweetness, by convention
bitterness, but in reality there are only atoms and space'}. In the above 
statement and a similar one by Kanad in Vaishyashik Sutras, there was only
a conviction that all the observed properties of things around us are
result of how the 'atoms' (the smallest, indivisible part of matter)
are put together. This was postulated without any idea of what the 
atoms were and/or how they are to be put together. It was a philosophical
statement. It took Sir Issac Newton, the father of Physics as we know it, to 
tell us how we should go about substantiating this and finding these 
atoms. He said in his Optics, '{\it Now the smallest Particles of 
matter may cohere by the strongest attractions and compose bigger particles of
weaker virtue....There are therefore agents in nature able to make particles
of Bodies stick together by very strong attractions and it is the Business of
experimental Philosophy to find them out}'. Thus 'experimental Philosophy'
is the earliest name one could give to this branch of science which dealt with
the issue.  Issac Newton was also the first one to put forward the 
theory of 'action at a distance' which explained the proverbial falling 
down of the apple, earth's going around the sun  and the strange appearance
of comets in the sky from time to time, in terms of  
{\it the same} interaction; viz. Gravitation.  At the time of Newton and 
for quite some time after that, thermodynamics might have been termed as the
branch of science that dealt with the structure of matter, as one could
describe the behavior of three states of matter; the solid, the liquid 
and the gas, in terms of  the laws of thermodynamics. However, already at this
time a further classification was known. Chemists already knew that one can 
classify objects by some properties which they seem to retain, independent 
of the state of matter: gaseous, liquid or solid. So Chemistry,  the study
of these chemical ``elements'', could have been  considered to be the branch of 
science dealing with this issue at that time.  The regularity of patterns
observed in masses, ionisation of various compounds, elements etc had led to 
the idea of ``atoms''.  The ordering of these chemical elements in the 
Periodic Table according to their properties, put forward by Mendeleev 
in 1876, is one of the earliest examples of recognising order/patterns
which can then be used as an experimental indication of the presence of 
underlying constituents. This phenomenon of  an observed 
order/pattern/regularity in the properties of `elemental' objects, being 
a smoking gun signal of a possible underlying structure, was to repeat oft 
in the years to follow.  

The modern saga of atomicity, after the Greeks/Vedantas, begins with Dalton at 
the end of 18$^{\rm th}$ Century. He observed that the chemical elements 
always combined in the same ratio to make a given compound. He postulated 
therefore that the chemical elements were made of units, which he termed 
``atoms''\footnote{By necessity
his ``atoms'' were essentially what we call ``molecules'' today.}. 
Avagadro further
found that all the gases combine in definite proportion of volume. That is the
number of molecules in a given volume, at a given temperature are the same.
This then led to the determination of molecular weights, molecular formulae 
and hence also of atomic weights. So already, by the early 19$^{th}$ century 
the Chemists
knew that all the molecular weights were rough integral multiples of that of
the hydrogen. So it was likely that the Hydrogen atom was the basic unit of 
them all! Thus the order in the atomic and molecular weights gave an indication
of the possible existence of a  basic building block in terms of the Hydrogen 
atom. Chemists 
kept on using the ``atomic'' theory without believing in the existence of the 
atoms till the advent of the kinetic theory of gases which gave a first 
principle derivation of all the observed property of gases\footnote{It is 
interesting to note that Einstein's famous first work in physics in 1905 on 
Brownian Motion was fueled by a wish to provide 'direct' evidence for 'atoms' 
to the straggling nonbelievers.}. The importance of the idea of atomicity to
the world of science is very graphically  expressed in the words of one of the 
greatest minds in physics of the 20$^{\rm th}$ century, arguably next only 
to Einstein, Richard Feynman. He says, '{\it If all the scientific knowledge 
in the world were to be destroyed and I can choose only one piece of 
understanding to be passed on
to future, I would choose to pass on the message that matter is composed of
atoms, ceaselessly moving and bouncing against each other}'. So at this point
in the human history, one could have said that the scientists had found the
``atoms'' which the Greeks /Kanad had postulated and which Newton had exhorted
the practitioners of ``experimental Philosophy'' to find.

Then in the later half of the nineteenth century came the discovery  which 
effectively defined the shape of physical and chemical sciences for the next 
century: the discovery by the great Michael 
Faraday that the electricity too comes in multiples of a basic unit. This and 
the experiments J.J. Thompson  performed  with the Cathode Rays, helped him 
discover the electron, the first elementary particle, in 1897. The world of 
particle physics
was born then. Indeed, Thompson went on to claim boldly '{\it Cathode rays are
matter in a new state, a state in which the subdivision of matter is carried 
much further than in the normal gaseous state, a state in which {\bf all} 
matter, - that is matter derived from different sources such as Oxygen, Hydrogen
etc. -  is one and the same kind, the matter being the substance from which all
chemical elements are built up.}'\footnote{To be honest this was a very
bold speculation on part of Thompson, not quite justified by the results 
he had gotten then.} Thompson had thus split the ``atom'' by 
proclaiming that it was made up of electricity : positive and negative !
Existence of an electron with the mass to charge ratio as measured by Thompson 
was shown to explain fine details of the atomic spectral lines under the
effect of a magnetic field, as calculated using an idea by Lorenz and
measured by Zeeman. These experiments in 1899,
helped electron make a transition from being a ``mathematical entity'' to being
a ``physical reality''. As a matter of fact the last mentioned was an important
step so that people could believe in the existence of the electrons in reality,

In the above description of the discovery of the electron, we see at work, all 
the three basic processes which have helped the physicists to arrive at the 
current picture of the basic constituents of matter. These were 
\begin{itemize}
\item [1)] Observation by Faraday that the electricity comes in units, from 
patterns in ionisation,
\item [2)] The experiments made by Thompson which showed him that the Cathode 
rays behave under the action of electric and magnetic fields as though they 
consisted of  particles with a ratio of charge to mass (the famous $e/m$) 
quite different from the Hydrogen ion,
\item [3)]The  measurement by Zeeman of the splitting of the atomic spectral lines
in a magnetic field and finding a value in agreement with that predicted using
ideas by Lorenz, assuming that a particle with that value of $e/m$ exists 
inside the atom.
\end{itemize}

In achieving the last it was necessary to have  an understanding of how 
to describe mathematically the interactions of the electron (a charged 
particle) with electromagnetic fields. This was in place by then, thanks
to Faraday and Maxwell.  This is an example of the synergy mentioned in 
the beginning of the article, 
between discovering what lies at the heart of the matter and figuring out the 
correct mathematical framework to describe interactions among the constituents 
of matter. 

As a matter of fact Thompson did not stop at making the bold speculation that 
matter was made of electrons, but gave a specific model for the Hydrogen atom
called the ``plum pudding model'' and had worked out in detail how the very 
``light'' electrons\footnote{Electrons were found to be about 1800 times lighter
than the Hydrogen atom.} could make up the atom. Then came another discovery
that shaped our thinking about matter again for decades to come: the famous 
Rutherford scattering experiment which many of us study in physics in the last 
years of our school these days. As a matter of fact, this is a classic 
example of one of the two paths to  ``elementarity'' that has been followed 
by scientists in their pursuit of what lies at the heart of matter. 
It is complementary to the other path mentioned earlier, where one uses
indications given by observing the pattern and order in properties of 
``elemental'' objects. Very often progress in these two paths,
took place side by side! 
I will give specific examples of this synergy in the context of Nuclear 
Physics and Elementary Particle Physics as we go along.

\begin{figure}[ht]
\begin{center}
\vskip -.5in
\includegraphics[scale=0.5]{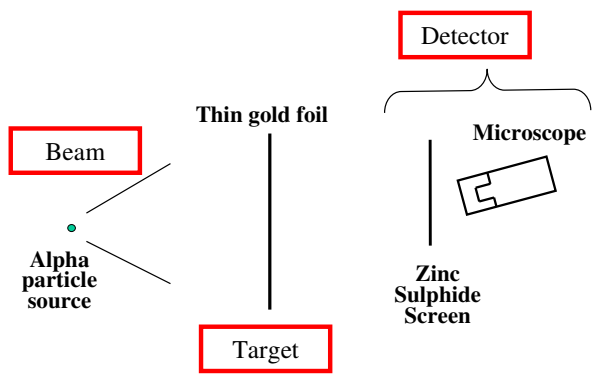}
\caption{\it Schematic depiction of Rutherford's experiment.}
\label{figru}
\end{center}
\end{figure}
Fig.~\ref{figru} depicts schematically the experiment conducted by Rutherford
in which he bombarded a thin gold foil with a beam of $\alpha$ particles 
emitted by the radioactive salts. The $\alpha$ particles carried a positive 
charge twice as much as the Hydrogen ion and weighed four times as much. He then
measured their angles of deflection after hitting the gold foil.
In technical terms he scattered  a {\it beam } of the $\alpha$ particles 
off the {\it target} of a gold foil and detected the scattered $\alpha$
particles in a {\it detector} made up of the zinc sulphide screen which 
produced scintillations when $\alpha$ particles hit it. Though it may not be 
very obvious at this stage, it was an attempt to  ``look'' inside the atom using
the $\alpha$ particles.  In this case the ``known'' piece of theory was the good
old Classical Mechanics started by Newton and  Electrodynamics : the theory
of how charged bodies moved under the action of electric and magnetic fields, 
honed to perfection by Faraday and Maxwell.
Using these two one could predict the trajectory of an $\alpha$ particle
passing through the distribution of the positive charge in the atom, which 
according to Thompson's Plum Pudding Model, was spread all over the atom 
uniformly with the $e'$s sticking up like plums. Recall that Thompson's 
experiment had revealed that the
$e/m$ ratio for the electron was much smaller than for the hydrogen ion
and thus the mass of the atom was expected to be concentrated in the positive 
charge. Thus it was expected that all the $\alpha$ particles will mainly feel
the positive charge and thus be deflected through very small angles. Imagine 
you are traversing through a 
big group of people uniformly spread over an area. You will have to change
your ``trajectory'' every so often as to avoid directly colliding with
another person and thus your trajectory will suffer small ``deflections''.
Now what Rutherford discovered was something exactly opposite to the 
expectations of Thompson's model. Most of the $\alpha$ particles traversed
through the foil without suffering any deflections at all, but those which
did deflect did so violently. Some of them even rebounded. In Rutherford's
own words, '{\it It was about as credible as if you had fired a fifteen inch
shell at a piece of tissue paper and it came back and hit you}'.  

This was a watershed of a discovery. Qualitatively it meant that all the mass
and the positive charge of an atom was concentrated in  a ``point''. 
So for most of the time $\alpha$ particles saw no charge which could repel them,
i.e. most of the atom was empty space. To carry the above mentioned analogy 
further, consider now that  you had to traverse the same road but the group of 
people now
had gathered in a tight crowd around some object of interest in the middle of 
the road. If you tried to pass through the crowd you will be pushed back, but 
if you were initially headed in the region away from this knot of people  you 
don't need to change your direction at all to pass through the road. This is
exactly what was being observed by Rutherford with his $\alpha$ particles and
the gold foil.  Of course, what fraction of time you will be repelled back will 
be decided by how widely spread the ``knot'' of people in the centre of the 
road is.  In more technical words, the fraction of $\alpha$ particles scattered
at a particular angle, called the angular distribution, can give information 
on the spatial extent of the charge distribution. Thus they can be used to 
``see'' whether the positive charge in the atom  had a structure or whether it 
was concentrated in a point. Rutherford showed that the angular distribution
observed by him, agreed with the one calculated using electromagnetism and 
Newton's laws of motion, assuming the positive charge  to be 
a ``point'' particle. He termed this point to be the ``nucleus'' of an atom. 
Thus now the next step in revelation of the structure of matter was taken:
the nucleus had arrived. The ``atom'' has been truly split and shown to consist
of a point nucleus and a whole lot of empty space containing the electrons. 
At this point the fundamental constituents of matter
were nuclei and electrons and they made up atoms, which in turn made up the 
molecules and so on. 

The next decades then saw further progress in the understanding of atoms in 
terms of a central nucleus and electrons as well as in 
that of the nuclei themselves. Atomic Physics and Nuclear Physics could then
have been said to be the branch of Physics dealing with the fundamental
constituents of matter around us. Emergence of patterns in the properties of 
nuclei, such as their masses, the spin angular momenta they carried, already 
indicated that the nuclei, though seen to be point-like, a-la Rutherford's 
measurements, perhaps had constituents. Note that while finite size of an 
object indicates that it has constituents, just because a particular object has
a size smaller than the least count of our best measuring stick, we can not
automatically conclude  that the object may not have constituents. All the 
observed regularities in the properties of the nuclei could be explained
by assuming that they were made up of protons (the hydrogen nucleus), 
and neutrons, neutron being  an electrically neutral particle with the same 
mass as that of the proton. Thus the list of the ``fundamental'' objects at 
this point would have contained only a few ``particles'': the photon $\gamma$ 
whose existence was deduced from ``Photoelectric Effect'' by Einstein, the 
electron $e$, the proton(p) and the neutron (n). To jump a few years in this
so far chronological narration, we could also include in addition the small 
``neutron'', {\bf i.e., the neutrino} $\nu$ that  Wolfgang Pauli, another 
intellectual giant, had had to reluctantly postulate to reconcile 
with the conservation of energy, linear momentum and angular momentum, the
experimentally observed properties of the $\beta$ particles observed in 
the decay of the  radioactive  nuclei. 

Let us note, as an aside, that perhaps this was one of the early examples 
where the requirement of such conservation principles, indicative of 
symmetries of the fundamental processes of physics, were used to postulate a 
new particle.  In this case the symmetry was the fact that the laws of physics 
are unchanged if we shift the origin of our coordinate system by a constant 
amount, rotate the coordinate axes and/or go to a frame of coordinates in a 
state of uniform relative motion.  Note that this is yet another way in which 
some of the fundamental constituents of matter announced their presence. We 
will have some more examples of the same later.

\section*{4. The tale of Nucleons}
But frankly this was just the tip of an iceberg. The heady developments of the 
early part of the 20$^{\rm th}$ century on the theoretical fronts, some of which 
were arrived at in an attempt to describe mathematically how the nuclei and the
$e'$s are held together in atoms and some which sprang from the genius of one 
mind (that of Albert Einstein), changed the way we thought about mechanics, 
space and time. Till then our ideas about these were solidly grounded in the 
laws laid down by Newton and Galileo. A quantum world in which time was no 
longer absolute was born. This is not the place to sketch out the developments 
of these desperately exciting times for physics which saw the emergence of 
Quantum Mechanics and the Theory of Relativity. We will, however, make use of
one very important concept of these times, that of the wave-particle duality
to take further this story of the hunt for the constituents of matter.

The question of whether a beam of light was made of corpuscles as Newton 
called them or whether it was a wave, was a topic of very hot discussions and
dissensions on the two sides of the English Channel. The issue  was decided in 
favor
of Huygens and the wave description of light by the late 19$^{\rm th}$ century.
However, Einstein's explanation of the Photo-Electric effect proved 
conclusively that light can be seen as made up of quanta of energies. Thus by
the early twentieth century the dual nature of light: as a wave as well as a 
particle, was an established fact. De Broglie hypothesized extension of this
dual existence to all the other particles such as the electron, postulating
that associated with a particle of momentum $p$, there is a characteristic 
wavelength 
\begin{equation} 
\lambda = \frac{h} {2 \pi p}, 
\label{eq1}
\end{equation}
where $h$ is  Planck's constant given by $6.6256 \times 10^{-34}$ joule-sec. 
At distance scales much larger than $\lambda$  we see particle behavior and 
at distance scales comparable or smaller than $\lambda$ the experiments will 
notice evidence of wave like behavior.  This was 
indeed verified in a famous experiment by J.J. Thompson's son G. J. Thompson. 

The above has a very interesting implication for the search of what lies at the
heart of matter. I mentioned two ways of doing this search, that have been used 
historically. The first being the use of patterns/regularities to learn about 
possible constituents and second being scattering experiments of Rutherford. 
However, so far I made no mention of the much more rudimentary ways such as
\begin{itemize}
\item 
[1)] breaking the system into its constituents by supplying enough energy: 
electric dissociation of molecules, photoelectric effect being a few examples 
\item 
[2)] using microscopes with better and better resolving power. This is what 
was done with biological systems helping us to arrive at the cellular theory 
of organisms. 
\end{itemize}

In fact, the experiments of the type performed by Rutherford are but a logical 
extension of this ``visual'' process mentioned in (2) above, just with a 
``microscope'' of higher 
resolving power. To understand this let us recall that if we want to decrease 
the minimum distance between two points up to which they can be told apart,
we need to use light of shorter and shorter wavelength. The phenomenon of 
diffraction, or bending of light around obstacles, is used to measure 
shorter and shorter distance scales and the limiting value is then the 
wavelength of the radiation used. The wavelength of visible light is several 
thousand Angstroms (one Angstrom is 100 million$^{\rm th}$ of a cm.). Using 
$X$--rays, which have a wavelength of a few Angstroms, we can measure distance
scales of the order of a few Angstroms such as distance between atoms in a
molecule etc.
Note however, that this ``seeing'' is no longer strictly visual.  The above 
wave-particle duality means that we can probe shorter and shorter wavelengths 
by replacing light with beams of accelerated particles. The higher the
energy (and hence the momentum) the shorter is the distance which we can
probe. This thus is  the genesis of electron microscopy. 

Again, let me make a small digression and once again jump ahead a few decades. 
Electron Microscopy has proved a very useful tool indeed once we learned
how to accelerate electrons to high energies. It has now been more than
50 years since  the first image of an Atom that was taken with a
Field Ion Microscope. In 1956, E. Mueller (an Indian physicist Dr. Bahadur was 
crucially involved in this exercise) presented  the image of a tungsten tip 
showing individual atoms. This is shown 
\begin{figure}[ht]
\begin{center}
\includegraphics*[scale=0.5]{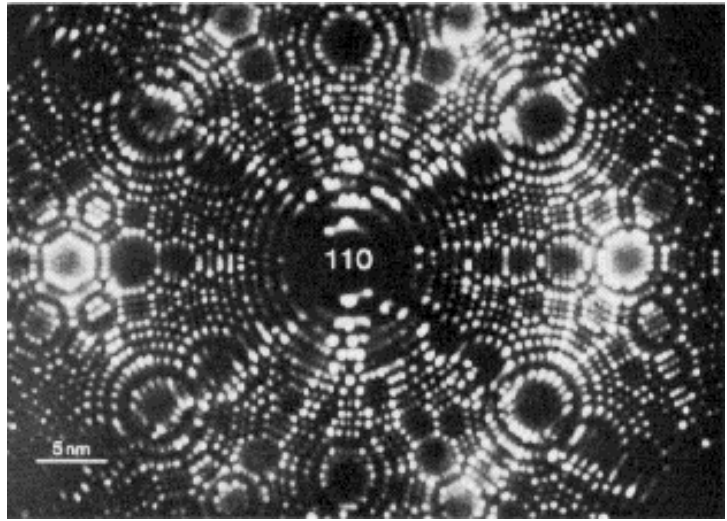}
\caption{\it Image of an atom.}
\label{figtip}
\end{center}
\end{figure}
in the Fig.~\ref{figtip}. Sure enough, by 1956 we knew atoms existed, we knew 
what their sizes were and so on, without ever having ``sighted'' the atom 
itself.  Still this achievement was a milestone in itself as it provided
the direct image of the atom and also because this marks the end of 
``visual'' sighting of constituents of matter. 

Let us get back to Rutherford and his scattering experiment. In his experiment,
he was using a beam of $\alpha$ particles which were being emitted by the
radioactive nuclei.  These had revealed that the atom was not ``indivisible'' 
and consisted of a ``point'' nucleus and electrons around it. However, further 
studies in Nuclear Physics had revealed that the nuclei themselves must be 
made up of protons and neutrons, bound together by an attractive force. 
If they were indeed made up of constituents why is it that Rutherford's 
experiment ``saw'' them as a ``point''? We can answer this question by
looking at the energies of the $\alpha$ particles used and their ``resolving'' 
power. These had energies of the order of  MeV and hence the corresponding 
wavelength given by Eq.~\ref{eq1} was about one tenth of a billionth of a cm.
or about a 100$^{\rm th} $ of an Angstrom. Thus it could resolve an object 
as not being a point, only if it was bigger than this distance. So 
Rutherford's experiment simply meant that the nuclei, if they had a size, 
were smaller than this.
On the other hand, since this wavelength is much smaller than an Angstrom, 
which is roughly the size of (say) Hydrogen atom, this beam was capable of 
revealing that the atom was not a point particle, but had a distribution where
the positive charge was concentrated in a very small region of the atom.

You see that we can thus use the high energy particle beams as a meter stick
to measure the size of an object, by scattering the beam off the object. The 
De-Broglie wavelength defined by Eq.~\ref{eq1} gives the limit of the length
scale which such scattering can probe. These high energy scattering experiments
thus are, but an extension of the process of trying to put an object under 
microscope to determine its structure and seeing its constituents. Of course,
the information is indirect and it is necessary to know the laws that govern 
the interaction between the ``probe'' and the ``target'' to be able to convert 
the
observed results into an information on the ``size'' of the object. Thus a 
knowledge of the laws of dynamics at level ``n'' is necessary to probe the
structure at level ``n + 1''.  This point can not be overemphasized. 
Since one needs beams of higher and higher energies to probe smaller and
smaller distance scales for existence of structure and/or constituents, the
subject of `elementary particle physics' is sometimes also called the 
subject of `high energy physics'. The tools we use to measure sizes of
objects changes with the size that they have! 
\begin{figure}[ht]
\begin{center}
\includegraphics*[scale=0.5]{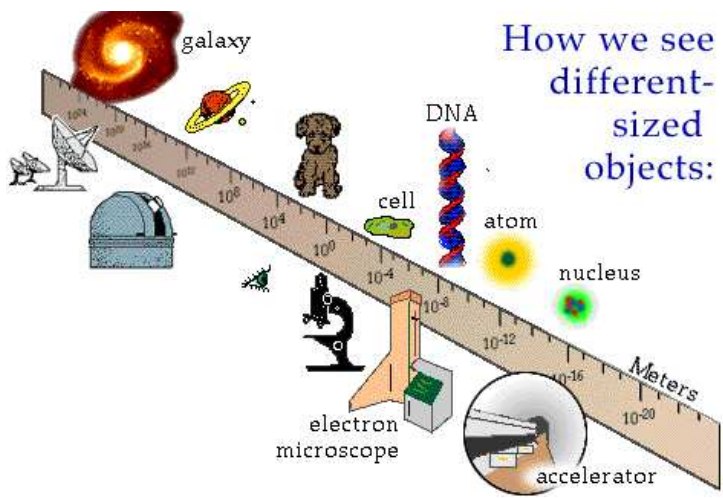}
\caption{\it Tools for `seeing' the objects and measuring their sizes.}
\label{figmeter}
\end{center}
\end{figure}
This fact is illustrated in Fig.~\ref{figmeter}. 

\subsection*{Unveiling the finite size of the nucleus}
The discussion so far about the processes which have unveiled the structure of 
matter tells us that this search proceeds essentially through three steps:
\begin{enumerate}
\item{}Seek the regularities/patterns in properties such as masses, spins etc. 
Very often these reflect {\it possible} existence of a more basic fundamental
units which makes the whole: an example would be atomic theory.
\item{} Measure the ``size'' of the constituents, which at the level of atomic
distances and smaller, is simply doing scattering experiments  using beams of 
higher energy particles to get probes of shorter and shorter wavelengths: 
example at the atomic level of this is Rutherford's experiment
\item{} A parallel and necessary step is also the development of a theory of
the dynamics that holds these units together. See if the observed properties
of the composites agree with the predictions of the theory: again at the 
atomic level the constituents revealed are nuclei and electrons, the 
subject dealing with the dynamics is Atomic Physics.
\end{enumerate}

\begin{figure}[ht]
\begin{center}
\includegraphics*[scale=0.5]{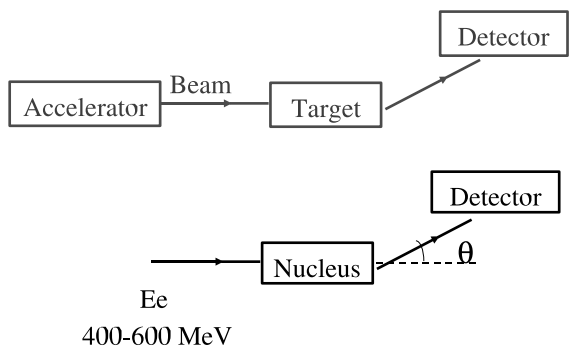}
\caption{\it  Nuclear/proton  analog of Rutherford experiment}
\label{figslac}
\end{center}
\end{figure}

Discussions of the earlier sections show that at the next level, in case of 
nuclei, the first step of indirectly inferring existence of the constituents 
of nuclei, had happened in the study of Nuclear Physics.  The next question
was  two fold: can one measure the size of the nucleus and can then
one ``see'' the constituents in the scattering experiments. Hence one had to 
devise an analog of Rutherford's scattering experiment, but one capable of
``resolving'' the nucleus beyond the limit of a hundredth of 
an Angstrom that Rutherford's experiments could put on its size. Clearly 
Rutherford knew the importance of higher energies already
when he said '{\it It has been long been my ambition to have available
a copious supply of atoms and electrons which will have energies transcending
those of the $\alpha, \beta$ particles.}' This became possible with the
advent of particle accelerators. Fig.~\ref{figslac} shows actually how such 
scattering experiments  were performed at the Stanford Linear Accelerator Centre which
began by probing how big the nuclei were. In the now famous experiments
by Hofstadter, electrons accelerated to energies of $400$--$600$ million 
electron volts were scattered  from nuclear targets. Note the 
similarity of  the beam-target-detector arrangement with the Rutherford case.
The wavelength of these electrons, $\lambda_e$, was about a 1000-10,000 times 
smaller than that of the $\alpha$ particles used by Rutherford. Again, all 
that the experimentalists did was to count the number of electrons scattered
at an angle $\theta$ as shown in the figure and compared it with the number 
expected for a point-like nucleus. With the arguments made above, it is clear 
that this ratio will be close to $1$ as long as $\lambda_e \gg R_{nucleus}$
and will start differing from $1$ as soon as the $\lambda_e \sim R_{nucleus}$.
Here the nucleus is assumed to be a sphere with radius $R_{nucleus}$. Indeed,
it is possible to study this ratio as a function of scattering angle $\theta$
and determine how nuclear charge is distributed in space. These experiments 
indicated that nuclei were about 10,000 -- 100,000 times smaller than atoms.
Mind you these experiments only proved that the nucleus has an extension in 
space, but could tell nothing whether it had any constituents. Of course since
the existence of neutrons/protons, called collectively a nucleon, was already 
inferred from studies in Nuclear Physics, the fact that the $p$ is not 
point-like  did not come as a big surprise.  The results of the earlier
scattering experiments which had not seen any indication of the presence of 
nucleons inside the nucleus, could be interpreted by 
saying that wavelength $\lambda_e$ was still much bigger than the separation
between the nucleons within the nucleus and hence it could not be resolved.
So at the end of this round of experiments, in 1960 or thereabouts 
(Hofstadter was awarded Nobel Prize in Physics for these experiments in 1961), 
\begin{itemize}
\item[1] The fundamental constituents of matter would have been $n,p,\gamma,e$ 
and the neutrino-$\nu$ whose existence was postulated by Pauli and confirmed in experiments
in Nuclear Physics as well as their anti-particles\footnote{Dirac's demand that 
the laws of Quantum Mechanics should look the same for all observers who are in 
state of uniform relative motion had predicted existence of an anti-particle 
for every particle, which would have the same mass, spin but opposite charge. 
The discovery of an anti-proton and existence of positrons had confirmed this 
prediction.},
\item[2]The elemental block of the earlier atomic level, the nuclei, were shown to 
have finite size and the sizes were measured by the scattering experiments,
\item[3] Nuclear Physics as a discipline had been able to give a good account of all
the observed nuclear properties by looking at nuclei as composites of the nucleons.
The dynamics of interaction between the nucleons was developed and studied by 
Nuclear Physicists.
\end{itemize}

The similarity of this description with the corresponding one presented above 
for the atomic case can hardly be missed.

\section* {5. The last layer?}
However, for various reasons none of the physicists around that time would have
agreed with the above list of particles as the list of the  fundamental 
constituents of matter. The first and  the foremost reason for this, was 
the observation of a very large number of particles similar to $n,p$, but 
somewhat heavier than them in the Cosmic Ray experiments. These experiments 
studied interactions of very high energy Cosmic radiation impinging on the 
atmosphere, producing large number of particles. The observation created a 
suspicion that may be $p,n$ are not fundamental after all. Another important 
indication that $p/n$ are not point-like and are a charge/mass  distribution, 
came from the observation that the  neutral neutron had a magnetic moment. 
According to Dirac's equation, mentioned already in the context of 
anti-particle prediction, the  neutral $n$ should have had no magnetic moment 
at all. Even more 
interesting was the observation by Gell-Mann and Zweig that the  pattern and 
the regularity that was exhibited in the properties of these supposedly 
elementary particles, could be explained by postulating that they were made 
up of a smaller number of more fundamental particles called quarks. However, no 
one had till then been able to break up the protons and neutrons into quarks, 
as had been possible in case of nuclei in nuclear reactions and/or decays. 
So people suspected that quarks were not ``real'' entities, but some kind 
of mathematical abstraction. Worse, quarks were required to possess fractional
electric charges (one-third or two-third the charge of an electron),
something that should have been noticed if quarks could travel around
by themselves. Hence many people regarded quarks as abstract entities,
and the quark model as ``just mathematics'', not unlike the Chemists of 
the 19$^{\rm th}$ Century  who used Dalton's atoms as a mathematical entity 
without believing in them.

According to the quark hypothesis, all the particles which experience
strong interactions are made up of quarks: a proton is a bound state of
two ``$u$-quarks'' and one ``$d$-quark'' and so on. The names $u$ and
$d$ stand for ``up'' and ``down'', but just like colour, these are
abstract concepts and could easily be given any other names. Indeed,
the attribute of being ``up'' or ``down'' is called ``flavour''.

To account for the particles then known, one required only three
different flavours of quarks: u(p), d(own) and s(trange).
A number of different high energy experiments gave results
consistent with the quark hypothesis. With the advent of higher
energies and the discovery of new particles, these three flavours
proved insufficient for the quark hypothesis to work. So three more flavours
were added: c(harm), b(eauty) and t(op). This (almost) accounts for the
quarks mentioned in Table 1.

\begin{figure}[ht]
\begin{center}
\includegraphics*[scale=0.6]{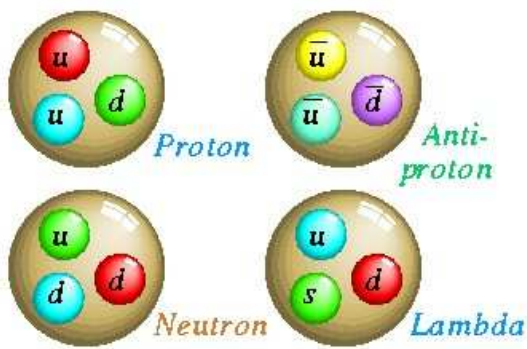}
\caption{\it How quarks and anti quarks of different types make up the strongly 
interacting particles.}
\label{quarkmodel}
\end{center}
\end{figure}

Today we believe that there exist precisely these six flavours of
quarks. Their presence is strongly, though indirectly, confirmed by
experiments, and is also required for consistency of the corresponding
theory, the Standard Model.  Fig.~\ref{quarkmodel} shows the composition 
of some of the known strongly interacting particles in terms of different 
quarks and (anti)quarks.

In 1965, soon after the original quark postulate, Greenberg, Han and
Nambu proposed that each flavour of quark comes in three different
species, differing only in an additional attribute which they called
``colour''.  They were led to this hypothesis by formal
considerations. Pauli's Exclusion Principle
tells us that the wave function of a collection of identical fermions
must be antisymmetric under the exchange of any two. Alternatively you
may know this as a statement that no electrons with the same energy
and spin can be in the same position.  However, the existence of a
particle called $\Delta^{++}$ posed a paradox for this principle. 
The paradox is straightforward to explain using ideas of  
quantum mechanics. 

The electric charge of $\Delta^{++}$ is 2 in units of electron charge,
and its spin is $\frac{3}{2}$ in units of $\hbar$. In terms of the
quark model, $\Delta^{++}$ must consist of three $u$ quarks. For it to
have a spin $\frac{3}{2}$, the spins of the three identical quarks
(each of spin $\frac{1}{2}$) have to be all aligned. Thus all the
quarks would be able to occupy the same position with the same spin
orientation.  More technically, this says that the net wave function
for $\Delta^{++}$ is {\it symmetric} under the exchange of any two $u$
quarks. That would contradict the exclusion principle, a fundamental
tenet of quantum mechanics. Thus the quark model, as understood at the
time, had to be wrong, or incomplete.

To resolve the paradox, Greenberg, Han and Nambu were led to introduce
an additional attribute, which they called ``colour'', taking three
different values (for example red, yellow and blue), solely so that
the wave function could be made antisymmetric under an exchange of
colour labels. In particular, the $\Delta^{++}$ would contain not
three identical $u$ quarks, but rather, one $u$ quark of each
colour. Then it would not be a problem to make the wave function
antisymmetric and save the exclusion principle.

A large number of measurements, such as the rate of decay of a neutral
pion into a pair of photons, gave evidence that the number of
quark species is really three times what was previously thought,
consistent with the colour hypothesis. 
However, at this time there was no evidence which would compel one to
accept quarks, ``colourful'' or otherwise, as genuine physical
entities. All attempts to observe spin
$\frac{1}{2}$ particles with fractional electromagnetic charges had
failed.  Thus, for a large class of physicists, the quark hypothesis
was just a kind of ``mathematics'' that explained very neatly a whole 
lot of observed
properties but did not require quarks to actually exist. 

In the meanwhile, indirect evidence for both the quark hypothesis as
well as the colour hypothesis was mounting, in different experiments
such as muon-antimuon pair production in pion-proton collisions,
or the production of strongly interacting particles in
electron-positron collisions.

One of the obvious things to do, as per the list given in the earlier 
section, was then to perform scattering experiments to see if indeed $p$ 
has a spatial extension to begin with. One could think later of addressing 
the question whether the scattering could reveal existence of these funny 
objects postulated from the requirements of patterns. As mentioned above, 
results of the various high energy experiments had agreed with the prediction 
of the ``quark'' model any way. So in that sense the third item on the ``to 
do'' list of the earlier section had been taken care of.

Similar to the experiments with the Nuclear targets, Hofstadter actually
confirmed that indeed the $p/n$ were charge distributions and the radius of this
distribution was 100,000 times smaller than one Angstrom : it was 
$\sim 1$ Fermi.
One thing to note here is that when we consider the scattering process,
\begin{equation}
e(E_e) + p \rightarrow e (E'_e) + p
\end{equation}
$e$ scattered at a given angle $\theta$ for a given energy of 
the incoming electron $E_e$ will have to have a given value of 
energy $E'_e$, (say $E'_0$).  The real surprise came 
as the energy of the electron was further increased to 
10,000 -- 20,000 million electron volts, reducing thereby the distance
it could probe  hundredfold compared to the size of the $p/n$. 
The scattered electrons at a given angle came with all possible energies, 
indicating thereby that may be the $p$ had something inside it. In principle
using the angle at which the electron travels  and its energy, one can back
calculate the momentum carried by whatever might be making up the proton. 
Thus the observed distribution in the energies of the scattered electron 
at a given angle then can thus be transformed into a distribution in momenta 
carried by these `constituents'. The most interesting observation was that
this distribution was the same when obtained using electrons of different
incident energies and scattered at different angles. Thus indeed, the
assumptions in the back calculations were correct and the electrons were 
bouncing off something else inside the proton.  Thus not only we knew that the
proton had some more things inside but we can also map the distribution
of the momentum of the proton that these constituents carried. The results 
indicated  that by now the wavelength of the probe was small enough to feel the effect of the individual scatterers inside the proton, separately.

Needless to say I have oversimplified this second coming of quarks. 
It suffices to say that the measurements of the above mentioned distribution 
in the energy of the scattered electrons, for a few different values of 
the scattering angles, allowed the physicists to even get information
about the possible spin as well as the electric charge of these elementary
constituents. It was indeed gratifying to see that these constituents seemed
to have all the properties (along with ``colour'') which they were required 
to have in the Quark Model. Thus one could identify these observed constituents 
of the $p$ with the quarks postulated by Gell-Mann and Zweig. 

As a matter of 
fact, results obtained by scattering higher and higher energy $e$ off the 
protons, indicated that the proton contains some other point like 
constituents to which the electron beam is blind, as they do not carry any 
electromagnetic charge. This was the first experimental glimpse of gluons.
Actually these scattering experiments, the so called Deep Inelastic Scattering
experiments, yielded very useful pointers which allowed physicists 
to formulate the right mathematical theory describing interactions of 
these quarks with each other and gluons. The Nobel Prize for Physics for 
the year 2004 was actually awarded for that theory called Quantum 
Chromo Dynamics (QCD).  But that can be a topic of a separate article.
The one feature of this theory that has implications for the 
present discussion is that, with increasing energy the number of 
constituents goes on increasing, since more and more quarks and gluons are
created inside the proton, when one tries to probe it with higher and higher
energy. {\bf That is, the increasing energies do not reveal any new 
constituents but reveal only this increasing number of quarks and gluons 
inside. This is in fact a firm  prediction of QCD.} In the simplified picture 
that I mentioned above, the peak in the scattered energy electron distribution 
will keep on shifting to values indicating an increasing number of 
constituents in the proton. Indeed such a rise was observed, precisely in 
the manner predicted by QCD, thus proving that electrons and quarks are indeed
point like and QCD the right framework to describe the dynamics of interactions
among quarks and gluons!

A reasonable question to ask is whether the existence of constituents inside 
a nucleus could also have been inferred from similar experiments with nuclear 
targets, in case we had not known about them before. The answer is yes. 
Such experiments were indeed performed and the results did indicate existence
of point-like scattering centers inside the nucleus just as in the case of the
proton and even the number of nucleons could be deduced. The only thing is that
the distance scale and hence the energies of incoming $e$ beams for which it 
was observed are scaled appropriately.

At present  experiments have been performed, not just with $e$ beams, but also 
$\mu$ beams and $\nu$ beams, with energies about 10-50 times higher than the
above. In an experiment in Germany,  30,000 MeV electrons are collided against 
protons which have an  energy of 920,000 MeV. This corresponds to using an 
electron beam with an energy of 100 Billion (1 Billion is one thousand million)
electron volts in the simple scattering experiment we have talked about. None 
of these experiments revealed any deviations from the expectations of the 
theory of point-like quarks and gluons, i.e., the above mentioned QCD. Thus 
there is no indication of any substructure of a quark up to a 1000$^{\rm th}$ 
Fermi. Thus we believe we have reached the end of the road in substructures.

So are we saying this simply because we don't have high enough energy probes? 
Indeed not.  This is where the part about the Particle Physics, the dynamics, 
which I have left out comes into play with full strength. Recall that this 
scattering (or equivalently ``seeing'') of the constituents was only {\it one} 
way in which we hunted for what lies at the  heart of the matter. At present 
every single piece of experimental observation agrees to a very high accuracy,
better than to one part in a 100 Millions at times, with the predictions of 
a theory which in these calculations, treats these quarks and leptons as 
point-like up to energies $\sim 10 $ billion billion eV. Thus we have 
an ``indirect'' {\it but very strong} proof that the quarks and the leptons 
are indeed point-like and have no further substructure.

It should be  added that I have sketched the path how we have arrived at the 
idea of quarks, in great detail and not said much about leptons. In fact they 
were not hunted for, but just came uninvited and made their appearance in the 
cosmic ray as well as  in the high energy experiments. Their properties never 
gave any indication of
substructure, the results of scattering reactions in which only leptons
were involved always agreed completely with predictions made
assuming that they were point-like. While, theory can not tell how many 
different repetitions of these pairs of quarks and leptons should be there, 
what the theory IS able to tell is that these should be equal in number. 
Indeed, this is satisfied  by the current list of the fundamental constituents
of matter. I have also not discussed  how the force carriers were 
``discovered''. But that requires a much more detailed discussion
of dynamics of the particle interactions, which we have left out.

Thus the discussion now clearly shows that the notion of what is elementary
is really decided by the resolving power of our probes, hence the distance 
scales we are interested in. All the discussion in these earlier sections
\begin{figure}[ht]
\begin{center}
\includegraphics[scale=0.5]{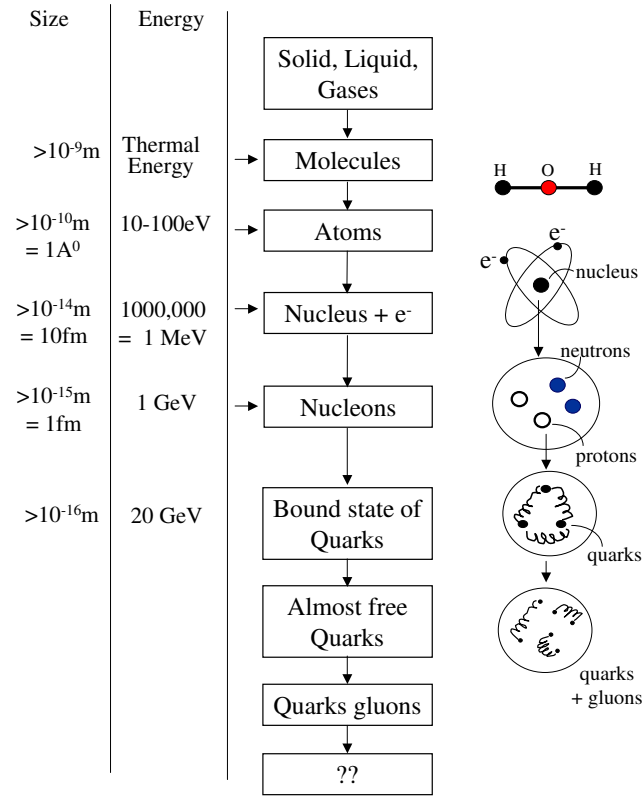}
\vskip -.5in
\caption{\it Constituents of matter at different distance and energy  scales}
\label{myfigscale}
\end{center}
\vskip -.45in
\end{figure}
can be summarised as shown in Fig.~\ref{myfigscale}. The figure shows the  
constituents of matter as we see them at different distance (and hence 
energy) scales.

\begin{figure}
\begin{center}
\includegraphics*[scale=0.4]{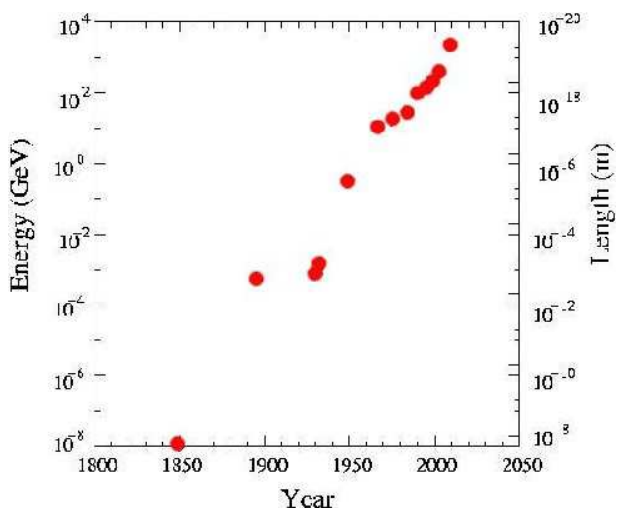}
\end{center}
\caption{\it How the energy frontier has moved in the decades.}
\label{fig:energyplot}
\end{figure}

Recall here also Fig.~\ref{figmeter}. This figure tells us
that high energy accelerators are our microscopes as we probe distance scales
of atoms/nuclei and further. Thus this journey into the 'Heart of Matter' is
accompanied by the development of accelerators. Figure~\ref{fig:energyplot}
shows the way the energy frontier has moved through
the decades and the distance scale of the new physics that this higher
energy has revealed. Through the early part of this journey the higher and
higher energy just revealed constituents  at smaller and smaller distance
scales. After the discovery of the  quarks lying at the heart of protons and
neutrons, the later increase in energy has brought about production of the
force carriers and helped develop/test the theory which can describe the
interactions among the fundamental constituents. The Large Hadron Collider
(LHC) that has just gone into operation at CERN in March 2010 and the
International Linear Collider (ILC) or CLIC that are under planning
are the spearheads of this energy frontier. We will discuss these next
and present what we expect them to achieve.

\section*{6. What Next?}
Following all the discussions in the earlier sections, one might be tempted
to ask,  now that  particle physicists seem to believe that they 
have arrived at a description of the ultimate constituents of matter and the  
interactions among them,  does it mean that this is the end of the road for 
the subject? Not at all. There are various reasons which  tell us that
we still have quite a way to go. 
\begin{itemize}
\item[1]
Firstly, the Higgs Boson which is predicted by these theories has to be found 
and shown to have exactly the properties that the theorists predict  it must 
have. This is almost like checking that the constituents 
of the $p$ as seen in the scattering experiments were indeed the quarks of 
the Quark Model. 

\item[2]Even if these 
experiments were to find this Higgs Boson there are still a lot of issues 
that need to be addressed and handled.  Even in the case of the Standard Model
itself, there are theoretical challenges 
such as understanding how mass less quarks, anti-quarks and gluons  make  
bound states that are massive, why free quarks never appear in nature etc. 
There are certain unsatisfactory theoretical issues  about the high energy 
behavior of the dynamical theories involving Higgs Bosons. Efforts to 
cure these problems have led to some popular extensions  beyond the SM. 
These predict existence of particles beyond what we have seen.

\item[3]
The $\nu$'s  have zero mass in the SM. However, the 
recent Nobel Prize winning experiments which showed that $\nu$ of one type can 
change into a $\nu$ of another type, have now firmly established that these
have a non-zero mass. Thus there are indications that the dynamics has 
something more than the SM. 

\item[4]
Further, even the three interactions that the SM 
addresses are not truly unified.
Particle physicists, including Einstein, have always held the dream of such
a unified description, one encompassing even gravity. So theorists are 
exploring ways to go beyond the dynamics contained in the SM. 

\item[5] In the heavy ion mode of the LHC the collisions can recreate energy
densities and temperatures which existed in the early Universe, giving us a
chance to actually study the transition of the ordinary matter into a Quark 
Gluon Plasma which again metamorphoses into hadrons. This part in the evolution
of the early Universe is opaque to various cosmological measurements and the
LHC is our only chance to study this in laboratory condition.
\end{itemize}

To summarise the above, we certainly need high energy accelerators which 
can give us direct evidence for the Higgs boson.  In addition, various 
extensions of the SM also make predictions of existence of new elementary 
particles and/or processes. The non-zero mass of the $\nu$s is indeed an 
extremely strong indicator for the existence of Physics beyond the SM. 
Studies of the $\nu$ sector may therefore provide us with theoretical 
and experimental clues to the Physics beyond the SM.  

It is obvious from the above discussions that the future of particle physics
rests on explorations on different fronts: a) theoretical investigations to
address various issues mentioned above and b) different experiments where 
these can be tested viz. These are i)experiments at high energy accelerators 
ii)experiments with high energy neutrinos and iii)the cosmological 
connections. In fact this state of affairs has been depicted very succinctly
\begin{figure}
\begin{center} \includegraphics*[scale=0.35]{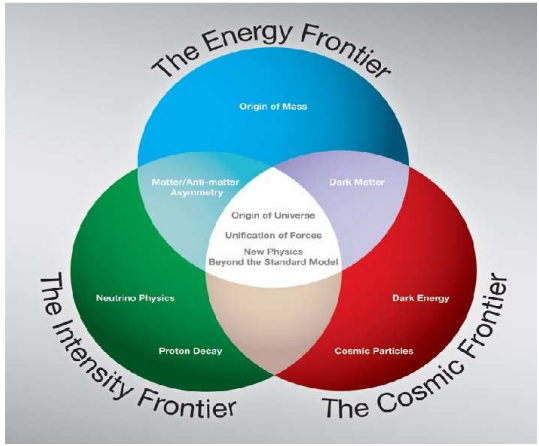} \end{center}
\caption{\it The three frontiers of progress in Particle Physics}
\label{fig:frontiers}
\end{figure}
in Fig.~\ref{fig:frontiers}, taken from the report of the High Energy
Physics and Astrophysics Panel (HEPAP), of the National Academy of Sciences, 
USA. The confluence of the results obtained at different frontiers will lead
to fundamental progress in our knowledge of the Universe. Indian Scientists
are in fact involved in activities on all the fronts.

On the energy frontier there is the Large Hadron Collider (LHC) which has gone
into action in March 2010, albeit with lower energy than was initially
foreseen; perhaps these teething problems remind us of the complexity of the
machine. The LHC is a proton-proton collider, where the two beams of protons
circulate in opposite directions in two beam pipes which run inside a tunnel
with periphery 27 km long. These two pipes intersect at a few chosen points 
so that the beams can collide. The beam bunches have to maintain their 
micrometer size diameter while traveling the distance of 27 km, which 
they traverse thousands of time. To achieve collisions of the required number of
high energy protons, the beams have to be steered by superconducting magnets
which are kept at a temperature of $1.9^{\circ}$ K. Building  such complex 
piece of machinery and making it work has been a matter of great joy and pride 
to the international high energy physics community. We can be very proud that
Indian engineers and accelerator physicists have been involved in building 
some part of this machine. The so called Precision Magnet Positioning Systems 
(PMPS) were manufactured in India. Not just this, Indian physicists have 
also been involved in building the mammoth detectors which are capable
of making very precise measurements (such as  determining the position of
a particle within a micrometer!) and thus can probe the mysteries of the 
laws of nature at their deepest level. India participates in the 
general purpose $pp$ detector CMS as well as the ALICE detector which
will study the heavy ion collisions.
\begin{figure}[htb]
\begin{center}
\includegraphics*[scale=0.5]{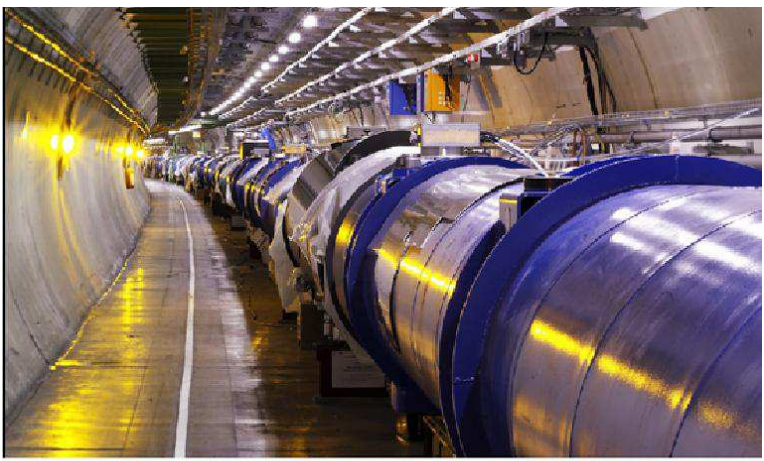}
\caption{\it The LHC tunnel with its accelerating magnets.}
\label{fig:lhctunnel}
\end{center}
\end{figure}
\begin{figure}[htb]
\begin{center}
\includegraphics[scale=0.3]{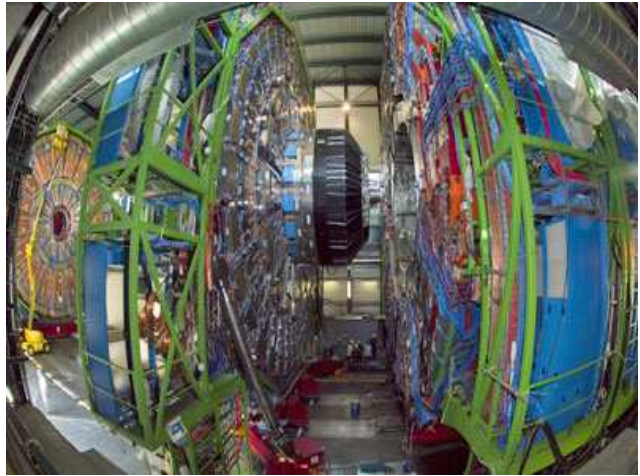}
\caption{\it The CMS detector in which India participates.}
\label{fig:lhccms}
\end{center}
\end{figure}
Figs.~\ref{fig:lhctunnel} and \ref{fig:lhccms}, show the LHC tunnel with 
the accelerating magnets and the cut out view of the CMS (Compact Muon
Solenoid) detector to which India has contributed. Thus the Indian scientific 
community is a part of this adventure.  Indian theorists are involved in 
the development of new and/or more refined theories of the Physics beyond 
the SM as well.  Indian physicists will also be involved in 
interpreting what the results coming out of LHC would mean for the SM 
and for the various theoretical ideas which go beyond the SM.

The international high energy physics community is convinced that it is 
necessary  to have an $e^+e^-$ collider, which should go in operation after LHC 
has run for a few years. This is truly an international effort in that even 
the optimal parameters for such a collider were decided by the entire 
international community. The same pattern continued in deciding
the optimal accelerator technology and now finally even the design of this 
accelerator is being done by an international team. Indian groups are part of
this global exercise as well and there exists an Indian Linear Collider Working
Group (ILCWG). A schematic drawing of the radio frequency cavities that would
have to be built, in order to construct the ILC is shown  
\begin{figure}[htb]
\begin{center}
\includegraphics[scale=0.4]{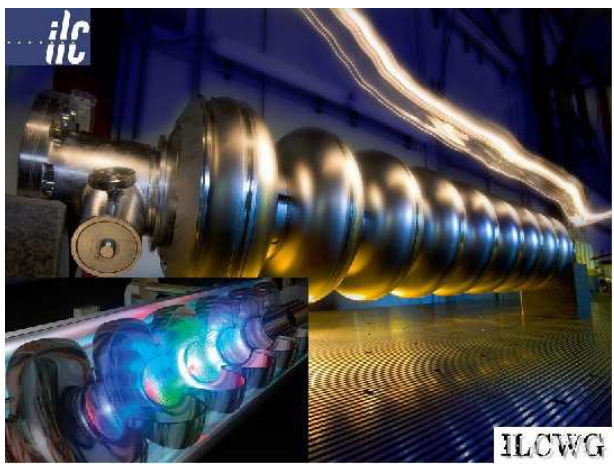}
\caption{\it Schematic drawing of a Radio Frequency cavity for the future
International Linear Collider.}
\label{fig:ILCdrawing}
\end{center}
\end{figure}
in Fig.~\ref{fig:ILCdrawing}
In fact the future consists not just of these collider experiments but also 
the gigantic Neutrino Experiments and India is part of that as well. Indian
High Energy Physicists are planning to build the Indian (International) 
Neutrino Observatory (INO). 
\begin{figure}[htb]
\begin{center}
\includegraphics*[scale=0.65]{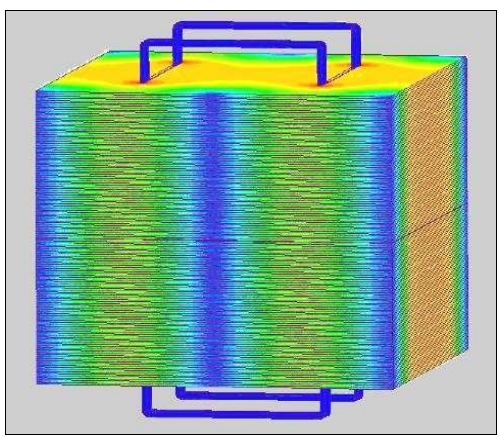}
\caption{\it Prototype of the INO iron calorimeter}
\label{fig:inocal}
\end{center}
\end{figure}
A prototype of the iron calorimeter they plan to use is shown in 
Fig.~\ref{fig:inocal}. You can get more information on the  ILC, INO etc. 
from the websites: {\it http://cts.iisc.ernet.in/Meetings/LCWG/index.html,
http://www.linearcollider.org/} and {\it http://imsc.res.in/~ino/}.

Actually, there is one more important laboratory where particle physicists can 
apply/test their theories and that is the Cosmos ! Cosmological observations 
now have  reached a degree of precision rivaling that of the HEP measurements.
Measurements by the Hubble telescope, the Sloan Digital Sky Survey, the 
Wilkinson Microwave Anisotropy Probe  etc., have now essentially tested 
the Standard Model of Big Bang Cosmology to a great degree and gone beyond it. 
Very high temperatures are supposed to have existed in the early Universe and 
at those temperatures all the fundamental particles would have existed. Their 
properties affect the evolution of the Universe in its first three minutes. The
number of mass less neutrino species, for example, affects what the value of 
the abundance of different type of elements in the Universe should be. Thus 
a knowledge of the spectrum of fundamental particles and their interactions 
is indispensable in the study of Cosmology. In the reverse, some of the ideas 
of physics beyond the SM can also be tested by their implication for cosmology. 
This interplay between High Energy Physics and Cosmology is most exciting. 

For example,  it is now well established that indeed the Universe consists of 
matter which does not shine, the so called Dark Matter (DM), whose presence
is revealed by its gravitational effects. At one stage $\nu'$s used to be a 
favorite DM candidate. However, very accurate measurements of the Microwave 
Background Radiation put now an upper limit on how much the $\nu'$s can 
contribute to the DM. 
One of the most promising ideas of going beyond the SM called Supersymmetry, 
necessarily predicts existence of a particle with exactly those  properties
that the cosmological calculations of fluctuations in the Microwave Background 
Radiation need. This particle called the lightest supersymmetric particle (LSP)
will be hunted for at the accelerators as well as in the  Astrophysical 
experiments. 

Further, it is found that matter  dominates over  the anti-matter in 
overwhelming proportions in the Universe, this is basically the reason why 
we exist.  In particular, the relative number density of $p/n$ (more generally 
baryon) with that of the $\gamma$'s is about 10 million times larger than the 
same for  $\bar p/\bar n$. This dominance can be understood in the Big Bang 
Model if  1) there existed interactions which did not conserve the 
proton number and 2)further violated the symmetry associated with the
combined transformation of charge conjugation (which exchanged particle
and antiparticle) (C) and space reflection (P), viz., CP.  Indeed, the 
grand-unified theories have ready-made candidates for interactions that 
violate the proton number. The quarks themselves do violate the CP symmetry 
by a very small amount. The quantitative calculation of this baryon 
asymmetry again seems to indicate that this observed and known CP 
violation present in the quark sector may not be quite enough, thereby 
indicating a possible class of models for going beyond the SM. Again, 
these models can be readily tested at the current and future colliders. 
Alternatively, same mechanism that gives masses to the $\nu$'s can also
give rise to adequate baryon asymmetry. This too can be tested in the 
accelerator experiments. Thus the Cosmology firstly provides strong 
constraints on the Particle Physics Models and secondly indicates regions of 
parameter space for these models where a satisfactory Baryon asymmetry can be 
obtained and thus makes the accelerator search for them more focused. This 
interplay between Cosmology and Particle Physics is truly fascinating.

The latest in the line is the so called Dark Energy. It seems to be proved 
that our Universe is slowly accelerating. This, along with the precision 
measurements of the Hubble constant and hence the age of the Universe, 
essentially imply that a large amount of Vacuum energy is present in the 
Universe. The answer to this issue may be linked with how we unify gravity 
with all the other interactions, what is the Quantum theory of Gravity etc. 
Since these are precisely the kind of issues that String Theorists are 
worrying about, it is likely that the latest Cosmological puzzle may find its 
solution in Particle Theory and Particle Theory 
may get hints about physics at the heart of matter through this. Again, 
only time can tell. But it is clear that our knowledge about the fundamental 
particles and their interactions can now address Cosmological issues. We 
in turn may get pointers for our searches of new physics and towards  our 
theories of the very fabric of Space and Time, from these Cosmological 
observations. So great things are in store.

\section*{Acknowledgments}
I wish to acknowledge a very enjoyable collaboration with Prof. 
Sunil Mukhi, on writing a  popular article on the Nobel Prize of 
2004, for the Science Education Magazine: Resonance.
I have freely borrowed from the above mentioned Resonance article
while writing the present one. I also
wish to acknowledge  support from the Department of Science and Technology,
India under grant number SR/S2/JCB-64/2007 (the J.C. Bose fellowship).

\end{document}